\documentclass[preprint,superscriptaddress,showpacs,preprintnumbers,amsmath,amssymb,aps,pre]{revtex4-2}
\usepackage[utf8]{inputenc} 
\usepackage{epsfig}
\usepackage{natbib}
\usepackage{booktabs}
\usepackage{amsmath}
\usepackage{times}
\usepackage{epstopdf} 
\usepackage{graphicx}
\usepackage{color}

\usepackage{afterpage}

\begin{document}

\title{Noise activates escapes in closed Hamiltonian systems}

\author{Alexandre R. Nieto}
\email[]{alexandre.rodriguez@urjc.es}
\affiliation{Nonlinear Dynamics, Chaos and Complex Systems Group, Departamento de
F\'{i}sica, Universidad Rey Juan Carlos, Tulip\'{a}n s/n, 28933 M\'{o}stoles, Madrid, Spain}

\author{Jes\'{u}s M. Seoane}
\affiliation{Nonlinear Dynamics, Chaos and Complex Systems Group, Departamento de
F\'{i}sica, Universidad Rey Juan Carlos, Tulip\'{a}n s/n, 28933 M\'{o}stoles, Madrid, Spain}

\author{Miguel A.F. Sanju\'{a}n}
\affiliation{Nonlinear Dynamics, Chaos and Complex Systems Group, Departamento de
F\'{i}sica, Universidad Rey Juan Carlos, Tulip\'{a}n s/n, 28933 M\'{o}stoles, Madrid, Spain}
\affiliation{Department of Applied Informatics, Kaunas University of Technology, Studentu 50-415, Kaunas LT-51368, Lithuania}

\date{\today}

\begin{abstract}

In this manuscript we show that a noise-activated escape phenomenon occurs in closed Hamiltonian systems. Due to the energy fluctuations generated 
by the noise, the isopotential curves open up and the particles can eventually escape in finite times. This drastic change in the dynamical behavior turns the bounded motion into a chaotic scattering problem. We analyze the escape dynamics by means of the average escape time, the probability basins and the average escape time distribution. We obtain that the main characteristics of the scattering are different from the case of noisy open Hamiltonian systems. In particular, the noise-enhanced trapping, which is ubiquitous in Hamiltonian systems, does not play the main role in the escapes. On the other hand, one of our main findings reveals a transition in the evolution of the average escape time insofar the noise is increased. This transition separates two different regimes characterized by different algebraic scaling laws. We provide strong numerical evidence to show that the complete destruction of the stickiness of the KAM islands is the key reason under the change in the scaling law. This research unlocks the possibility of modeling chaotic scattering problems by means of noisy closed Hamiltonian systems. For this reason, we expect potential application to several fields of physics such us celestial mechanics and astrophysics, among others.

\end{abstract}

\pacs{05.45.Ac,05.45.Df,05.45.Pq}
\maketitle
\newpage
\section{Introduction} \label{sec:Introduction}

Chaotic scattering is an important area of research in nonlinear dynamics 
due to its fundamental applications to a wide variety of fields such as physics \cite{Seoane13}, chemistry \cite{Ezra,Kawai}, medicine \cite{Schelin,Schelin2} and biology \cite{Scheuring,Tel}. In the context of physics, 
many chaotic scattering processes can be modeled by using open Hamiltonian systems. These systems are characterized by a potential that exhibits one or more exits through which the particles can escape to infinity after describing a transient chaotic motion. Some paradigmatic examples are the Hénon-Heiles Hamiltonian \cite{HH64}, with applications to astronomy and celestial mechanics \cite{NavarroAA}, and the Barbanis potential, used in quantum dynamics \cite{Babyuk} and chemical physics \cite{Sepulveda}. Certainly, most of the research concerning chaotic scattering in open Hamiltonian systems has been done in purely conservative cases \cite{Contopoulos90,Bleher,Barrio09}. 
However, many dynamical systems are not isolated from its environment or they have internal irregularities that can be considered in the dynamics by introducing external perturbations. For this reason, recently different lines of research have emerged, seeking to elucidate the effects of weak perturbations on the escape dynamics. In particular, the periodically forced \cite{Nieto18,Blesa14}, dissipative \cite{Seoane07,Motter} and noisy \cite{Altmann,Seoane08} cases have been studied.

Regarding the effects of noise, much work has been done in both chaotic maps and open Hamiltonian systems. In the case of chaotic maps, it has been shown that the noise can destroy the fine structure of the Kolmogorov-Arnold-Moser (KAM) islands \cite{Mills} and generate escapes \cite{Rodrigues,Silva}. In discrete systems even a very weak noise intensity turns the dynamical behavior hyperbolic and then the survival probability decays exponentially \cite{Rodrigues}. About open Hamiltonian systems, a deep analysis on the effects of the noise in the phase space structure and in the exit basins topology has been carried out both in the presence of additive noise \cite{Bernal13} and noisy periodic forcing \cite{Gan}. In this kind of systems, the decay law of the survival probability remains algebraic 
for low noise intensities \cite{Bernal13,Seoane09}, what implies that the 
dynamical behavior can remain nonhyperbolic under weak perturbations.

In addition to all the above, a surprising phenomenon called noise-enhanced trapping appears in both chaotic maps \cite{Altmann} and open Hamiltonian systems \cite{NietoNET}. The mechanism that generates this phenomenon is characterized by an increase on the average escape time for some critical value of the noise.

As we have seen, the noisy scattering problems have been attacked from many angles and a deep understanding of the main effects of the random perturbations in both discrete and continuous systems has been addressed. However, as far as we know, the influence of the noise in closed Hamiltonian 
systems has received less attention. Many examples of Hamiltonian systems, including the Hénon-Heiles system and the Barbanis potential mentioned above, exhibit an energy threshold $E_e$ that separates two very different regimes. Under $E_e$ the isopotential curves are closed, so the particles are bounded. On the contrary, for values over $E_e$ the potential well exhibits at least one exit through which the trajectories can escape to infinity after describing a transient chaotic motion. In this manuscript, we have 
included a source of uncorrelated Gaussian noise in closed Hamiltonian systems. Under the influence of random fluctuations the energy is not conserved and noise-activated escapes occur, turning the closed Hamiltonian system into a chaotic scattering problem. We have found that the evolution of the average escape time with the noise intensity follows a different algebraic scaling law in the case of low and high noise intensities. This is a consequence of the destruction of the stickiness \cite{AltmannS,Bunimovich} of the KAM islands. Moreover, we have obtained that the general effects of the noise in the system are different from the case of open Hamiltonian systems. In particular, the noise-enhanced trapping, relevant phenomenon in noisy open Hamiltonian systems, does not influence the dynamics when the conservative system is closed.

The structure of this paper is as follows. In Sec.~\ref{Model}, we
describe the model of closed Hamiltonian system that we have used in this 
work, the H\'{e}non-Heiles Hamiltonian. The qualitative explanation about 
the noise-activated escape phenomenon is carried out in Sec.~\ref{NAE}. In Sec.~\ref{ALS}, we show the result of the computation of the average escape time in function of the noise intensity, showing a transition region at which the scaling law changes. The reasons of the change in the regime when the noise is increased are explained in Sec.~\ref{PE} and Sec.~\ref{EF}. In the former we provide a strong numerical evidence that confirms the gradual reduction of the stickiness of the KAM islands, while in the later we bring some theoretical reasoning based on the energy fluctuations. Finally, in Sec.~\ref{Conc}, we present the main conclusions of this research.

\section{Model description}\label{Model}

To carry out this research about the effect of noise in closed Hamiltonian systems we have used as a model the well-known Hénon-Heiles Hamiltonian. Since it appeared in the literature in $1964$ as a model of a galactic potential, the system has been broadly studied \cite{Barrio08,Aguirre01,Vallejo03,Aguirre09} due to the vast assortment of dynamical behaviors that exhibits for different values of the energy. The system is characterized by a nonlinear potential with a triangular symmetry that, together with the kinetic energy, defines the Hamiltonian

\begin{equation} \label{eq:HH_Hamiltonian}
    {\cal{H}}=\frac{1}{2}(p_x^2+p_y^2)+\frac{1}{2}(x^2+y^2)+x^2y-\frac{1}{3}y^3,
\end{equation}
where $x$ and $y$ denote the spatial coordinates and $p_x$ and $p_y$ are the two components of the generalized momentum.

\begin{figure}[htp]
	\centering
	\includegraphics[width=0.5\textwidth,clip]{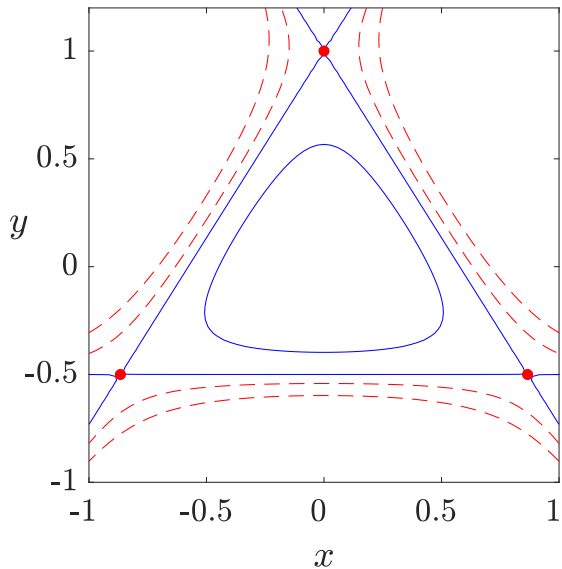}
	\caption{Isopotential curves of the Hénon-Heiles system for	different values of the potential $V (x, y) = \frac{1}{2}(x^2+y^2)+x^2y-\frac{1}{3}y^3$. The blue solid curves represent energy values $E=1/10<E_e$ and $E_e=1/6$, while the dashed red curves represent energies over the threshold $E_e$ ($E=0.2$ and $E=0.3$). The red dots are located in the saddle points that flank the exits. }
	\label{fig1}
\end{figure}

The system undergoes deep changes on its dynamical behavior for the threshold value of the energy $E=E_e$. This value is also called escape energy since it separates the open and closed regimes. If the energy is less than or equal to $E_e$ the isopotential curves are closed and the trajectories are bounded. On the contrary, if the energy is greater than $E_e$ the trajectories can move over the saddle points and hence the isopotential curves open up. In this situation three symmetrical exits separated by an angle of $2\pi/3$ radians emerge. To visualize the qualitatively different regimes and the system itself, we represent the isopotential curves for four different values of the energy in Fig.~\ref{fig1}.

Here, we consider that the system presents internal irregularities and/or is affected by its surrounding environment. To take these effects into consideration it is natural to introduce a periodic driving or a weak noise in the equations of motion \cite{Kandrup2000}. In this research, we consider the influence of a random force, that we idealize as an uncorrelated additive Gaussian noise with zero mean.

After this considerations, the motion is described by the following equations \cite{Seoane08}

\begin{equation} \label{eq:eq motion}
    \begin{aligned}
    	\dot{x} & = p_x, \\
    	\dot{y} & = p_y, \\
        \dot{p_x} & = -x - 2xy + \sqrt{2\xi}\eta_x(t), \\
        \dot{p_y} &= -y - x^2 + y^2 + \sqrt{2\xi}\eta_y(t),
    \end{aligned}
\end{equation}
where $\eta_x(t)$, $\eta_y(t)$ are Gaussian white noise processes $X_t\sim {\cal{N}}(0,2\xi)$ of intensity $\xi$ and autocorrelation function $ \langle \eta(t')\eta(t)\rangle=\sqrt{2\xi}\delta(t'-t)$.

As a means to solve the system of differential equations, we have implemented a stochastic second-order Heun method \cite{Kloeden}, that has been already tested and used in previous research concerning the noisy Hénon-Heiles system \cite{Seoane08,Bernal13}. We have tested the effectiveness of the method by ensuring the convergence of the average escape time when the integration step is reduced. After some testings, we have set the integration step in a very small value $h=0.001$.  We also want to highlight that all our simulations have been done after averaging an enormous number of trajectories, as it is reflected in the small values of the errors obtained in the numerical simulations.

The heavy computational load is one of the disadvantages of noisy Hamiltonian systems. In these systems the convergence of many magnitudes like the escape times is much slower than in conservative systems. Moreover, since the noise generates final state sensitivity \cite{NietoFSS}, to show some results it is necessary to compute the trajectory of the same initial condition many times until convergence. Therefore, for the sake of the computational efficiency, and without loss of generality, in all the simulations of this manuscript we have used the threshold value of the energy $E=E_e$. By reducing the energy, we should obtain qualitatively identical results but with a clearly higher computational cost.

\section{Noise activates escapes}\label{NAE}

\begin{figure}[htp]
	\centering
	\includegraphics[width=0.55\textwidth,clip]{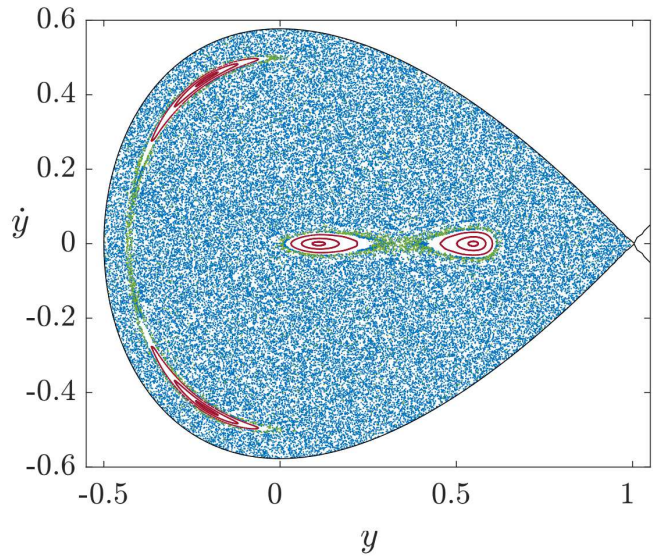}
	\caption{$x = 0$ Poincaré section in the $(y,\dot y)$ phase space 
for the Hénon-Heiles system with $E=1/6$. Here one chaotic trajectory (blue), one sticky chaotic trajectory (green) and three quasiperiodic 
orbits (red) are represented.}
	\label{Fig2}
\end{figure}

As we have mentioned in the Introduction, for energy values $E\leq 1/6$ the Hénon-Heiles is a closed Hamiltonian system. Hence, the trajectories are bounded inside the potential well. Even so, the dynamical behavior is not trivial at all since chaotic trajectories and quasiperiodic orbits coexist. Depending on the initial condition a trajectory can evolve in a different way, either describing a regular orbit inside a KAM torus or moving chaotically. Although the chaotic trajectories cannot enter inside the KAM islands, if they move close enough to one of them they become affected by its stickiness. This implies that the trajectories can 
spend an arbitrarily long time surrounding the neighborhood of the KAM islands before exploring the rest of the phase space. To visualize these three different orbits, we have represented the $x = 0$ Poincaré section in the $(y,\dot y)$ phase plane for the energy $E=1/6$ in Fig.~\ref{Fig2}. By looking at this picture, it is straightforward to find out the origin of the terms \textit{chaotic sea} and \textit{KAM island}. Following the analogy, we represent in blue the chaotic sea, while in red and green we depict three quasiperiodic orbits and a sticky chaotic trajectory, respectively.

When we consider that the system is affected by its environment in such a 
way that a Gaussian white noise is included in the equations of motion, the situation described above changes drastically. Since the noise affects the speed of the particles (and therefore the energy), the system is not conservative anymore. Due to the fluctuations in the energy, the system is trembling and a chain of positive fluctuations in the energy leads the particle to escape, even if the noise intensity is very low. As it has been shown in previous works, even feeble perturbations such as periodic forcing \cite{Nieto18}, dissipation \cite{Seoane07} and noise \cite{NietoNET} destroy the KAM islands structure, so particles are not expected to remain in the well. Despite the fact that the particles will surely escape from the scattering region, the main characteristics of the scattering process will be decided by the complicated effects of the random fluctuations on the chaotic and regular dynamics.

Because noise activates escapes, the closed Hamiltonian becomes a system with all the ingredients of a chaotic scattering problem. Therefore, the escape dynamics can be studied by means of the usual tools like the average escape times. Furthermore, recent concepts such as the probability basins \cite{NietoFSS} can be also employed.

\begin{figure}[htp]
	\centering
	\includegraphics[width=0.72\textwidth,clip]{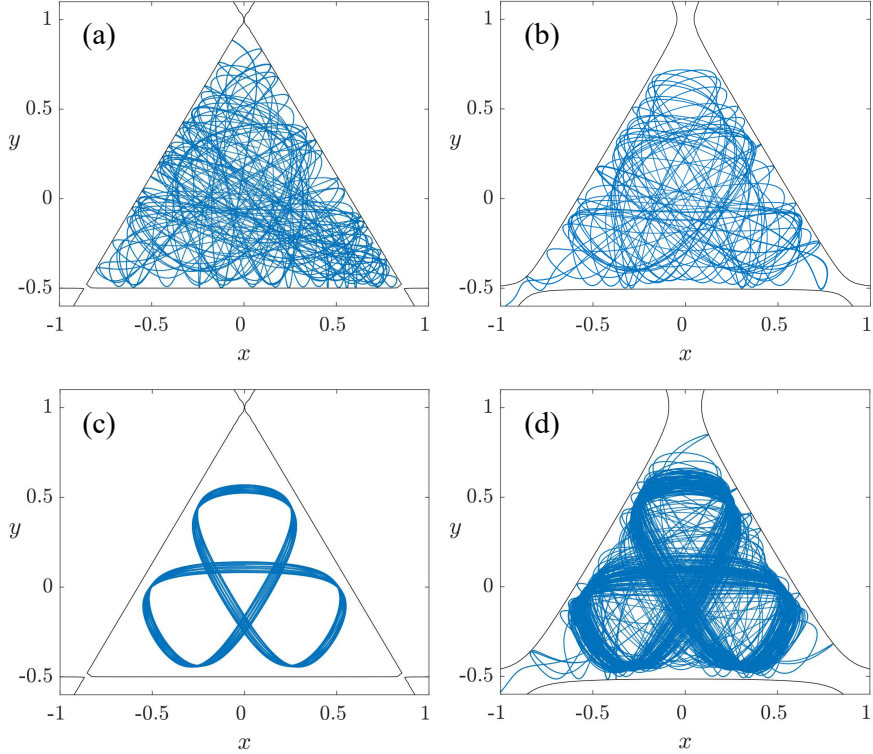}	
	\caption{Projection in the $(x,y)$ plane of (a) a chaotic trajectory and 
(c) a quasiperiodic orbit in the noiseless Hénon-Heiles system with $E=1/6$. Panels (b) and (d) represent trajectories with the same initial condition that we have used in panels (a) and (c), respectively, but 
in the presence of noise with intensity $\xi=10^{-7}$. Even if the noiseless system is closed, both particles escape after a finite time (b) $t=275$ and (d) $t=2214$. The isopotential curves in panels (b) and (d) correspond with the energy of the system in the moment that the particle escapes. }
	\label{Fig3}
\end{figure}

In order to visualize the noise-activated escape, we show in Fig.~\ref{Fig3} the plots corresponding to two initial conditions for noiseless case and when a noise of intensity $\xi=10^{-7}$ is present, what implies escaping. While the chaotic trajectory escapes after a short time $t=275$, the noise needs a clearly higher time $t=2214$ to activate the escape from the KAM island. Needless to say, if we repeat these two simulations with exactly the same initial conditions, the escape times and the exits could be different.

In order to choose the initial conditions, here and throughout the entire work, we have used the tangential shooting method \cite{Aguirre01}. Since the phase space is four-dimensional, it is necessary to find a suitable Poincaré section that does not lead to the loss of information about the system. The tangential method is quite common since it preserves the equiprobability of the exits and detects the main KAM islands. Following this method, once we set an initial position $(x_0,y_0)$ in the physical space, we launch the particles  in such a way that the velocity vector $\mathbf{v}$ is tangent to a circumference centered at the origin and passing through the point $(x_0, y_0 )$. Hence, the Poincaré section is defined by the set of points that verify $\mathbf{r} \cdot \mathbf{v}=0$, $\mathbf{r}$ being the vector that goes from $(0,0)$ to $(x_0,y_0)$. Since there are two antiparallel vectors that satisfy the previous condition, we arbitrarily set $\mathbf{r} \times \mathbf{v}>0$.

When observing Fig.~\ref{Fig3}, we note a counter-intuitive result since in the left panels the isopotential curves are closed, while in the right panels are open. At first glance, one might think that the simulations were carried out with different energy values $E=1/6$ and $E>1/6$. However, the opening of the potential well is a consequence of the presence of noise. Even if the noise has zero mean, the energy changes in time allowing the particles to evolve with higher or lower energies than in $t=0$. When the escape occurs, of course the energy is $E>1/6$, so if we represent the isopotential curve in the moment of the escape it will be open.

\begin{figure}[htp]
	\centering
	\includegraphics[width=0.55\textwidth,clip]{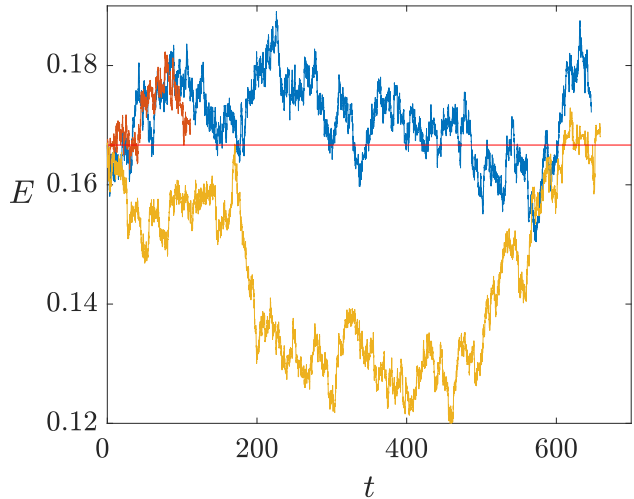}	
	\caption{Evolution of the energy of three simulations of the same initial condition with a noise intensity $\xi=10^{-6}$. The horizontal red line indicates the initial energy $E=1/6$. The trajectories can evolve with  
higher or lower energies than the original, but all of them escape from the scattering region after describing transients of different duration. }
	\label{Fig4}
\end{figure}

In order to bring a general view about the mechanism of the noise-activated escape, we represent the energy evolution of three launchings of the same initial condition $(x,y,p_x,p_y)=(0,0.4,-0.47,0)$ affected by a noise intensity $\xi=10^{-6}$ in Fig.~\ref{Fig4}. This initial condition would generate a bounded chaotic motion in the conservative system. As we can see, the yellow colored trajectory moves during almost all its transient with energies $E<1/6$, while in the blue colored trajectory the opposite occurs. Despite this different energy evolution, the escape time is similar in both cases. This is because the opening of the potential well does not lead to the immediate escape of the particles, but it is a result of the noisy chaotic scattering process. In particular, as it has been shown in recent research \cite{Altmann,NietoNET}, before escaping the particles can jump in and out of the KAM islands, generating a complicated evolution that involves both chaotic and regular motion. The red colored trajectory is another example of the complicated escaping mechanism. This trajectory follows initially a similar evolution to the blue colored one, reaching a maximum energy $E>0.18$. However, the escape occurs after decreasing its energy to values close to the threshold $E=1/6$. As we can see, the noise-activated escape is simply a consequence of the noisy nature of the system, but the escape dynamics is a complicated phenomenon in which the noise and both the chaotic and regular behaviors are involved.

\section{Scaling law of the average escape time}\label{ALS}

In the previous section we have explained in a qualitative manner the noise-activated escape phenomenon. In order to get a quantitative evidence of the influence of the noise intensity on the escape dynamics, in this section we compute the average escape time $\bar{T}$ in function of the noise intensity. The relation between these two variables is explained by a scaling law, that offers relevant information about the changes in the escape dynamics when the noise intensity is increased. Without further analysis we cannot state more about the evolution of $\bar{T}$ than that some limit behaviors should be satisfied. First, since for $\xi=0$ the system is closed, the limit
\begin{equation} \label{lim1}
	\lim_{\xi\to 0} \bar{T}=\infty
\end{equation}
must be fulfilled.

On the other hand, if $\xi\gg 0$ the noise effects dominate the dynamics and the trajectory immediately escapes, so
\begin{equation} \label{lim2}
 \bar{T}(\xi \gg 0)\approx 0
\end{equation}
is expected.

\begin{figure}[htp]
	\centering
	\includegraphics[width=0.5\textwidth,clip]{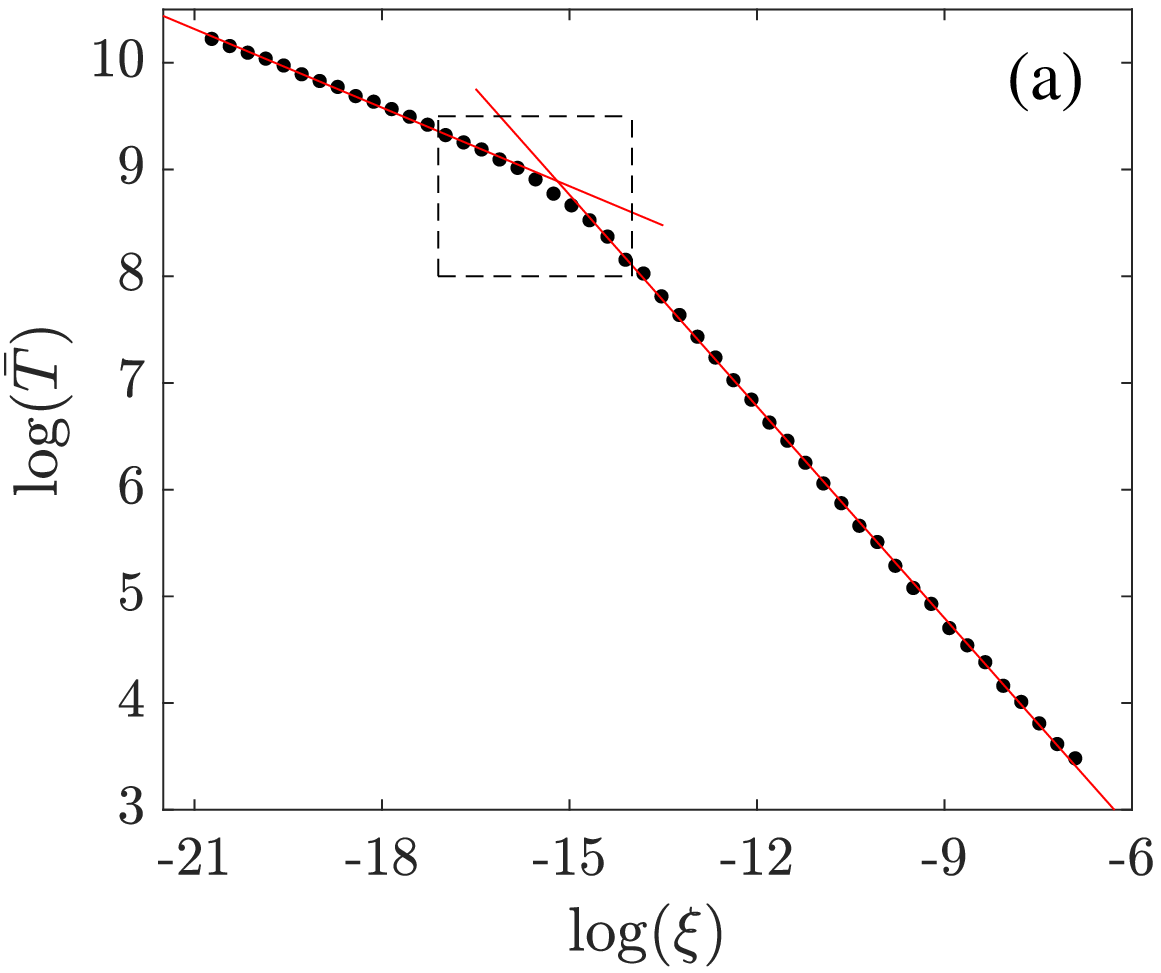}
	\includegraphics[width=0.5\textwidth,clip]{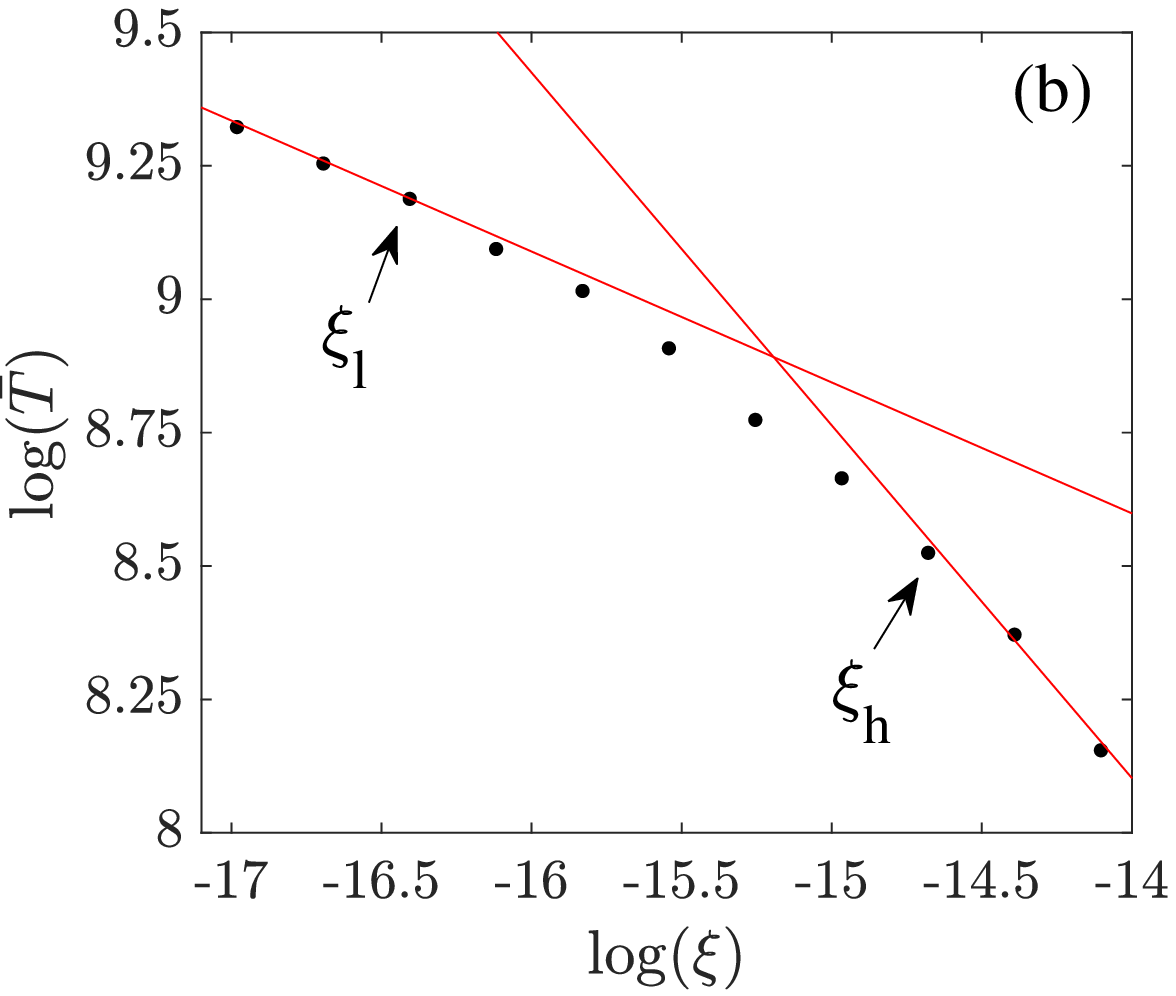}
	\caption{Logarithmic plot showing the evolution of the average escape time with increasing noise intensity. The red lines correspond to two different linear fits for low [$\log \bar{T} = -0.2454(\pm 0.0029)\log{\xi}+5.163(\pm 0.053)$] and high [$\log \bar{T} = -0.6611(\pm 0.0022 ) \log{\xi}-1.153(\pm 0.025)$] noise intensities. In both cases the linear correlation coefficient is very close to $-1$ ($r=-0.9989$ and $r=-0.99985$, respectively). We have not included error bars in the graphic since the errors are so small that they could not be appreciated. Panel (b) shows a magnification of the dashed region drawn in panel (a), which is centered on the transition region.}
	\label{fig5}
\end{figure}

In order to carry out the numerical simulations of the evolution of the average escape time, we have taken different values of the noise intensity in the range $\xi\in[10^{-9},10^{-3}]$. We have calculated the average escape time $N=30$ times for every noise intensity. We have computed the escape time of $10^4$ random initial conditions in the physical space for each one of these $N$ simulations. The final value $\bar{T}$ is the mean of the $N$ repetitions and we have considered its error, $\Delta \bar{T}$, as $3$ standard deviations (i.e. a $99.7\%$ confidence level). On the other hand, the error estimation of $\log(\bar{T})$ is $\Delta \log(\bar{T})=\partial \log(\bar{T})/ \partial \bar{T} = \Delta \bar{T}/\bar{T}$. We highlight the high precision of the results, the maximum error obtained being $\Delta \log(\bar{T})<0.03$. The result of this simulation is shown in Fig.~\ref{fig5}(a), where two different regimes can be observed, each one characterized by a different algebraic law. It is worth noting that the evolution of $\bar{T}$ with increasing $\xi$ cannot be explained by means of a single standard scaling law such as algebraic, logarithmic or exponential. For low noise intensities under $\xi_l\approx 1.5\times 10^{-7}$, an algebraic law $\bar{T}_l$ explains the evolution of the average escape time. On the other hand, for high noise intensities over the critical value $\xi_h\approx 4\times 10^{-7}$, another algebraic law $\bar{T}_h$ satisfies the numerical results. Between these two regimes a smooth transition region appear, and it is not characterized by either $\bar{T}_l$ or $\bar{T}_h$. This fact can be observed in Fig.~\ref{fig5}(b), where we represent a magnification of Fig.~\ref{fig5}(a) close to the transition region. In order to give an account of the algebraic relation between the variables, the results are represented in a log-log plot together with two accurate linear fits in Fig.~\ref{fig5}. By simply considering the slope and the $\ln{\bar{T}}$-intercept obtained by the least squares method, we can get the parameters of the underlying algebraic law
\begin{equation}\label{scaling}
	\bar{T} = \alpha\varepsilon^{-\beta},
\end{equation}
where $\alpha$ and $\beta$ are positive constants. Clearly, this equation 
satisfies both Eq.~(\ref{lim1}) and Eq.~(\ref{lim2}).

In the case of noise intensities $\xi<\xi_l$, the parameters are $\alpha_u=174.7(\pm 9.3)$ and $\beta_u = 0.2454(\pm 0.0029)$, while for noise intensities over $\xi_h$ we have obtained $\alpha_o=0.3157(\pm 0.0078)$ and $\beta_o = 0.6611(\pm 0.0022 )$. The errors have been considered as $\Delta \beta = \Delta a$ and $\Delta \alpha = e^b\Delta b$, $\Delta a$ and $\Delta b$ being the errors of the slope and the $\ln{\bar{T}}$-intercept, respectively.

The derivative of Eq.~(\ref{scaling}) indicates the reduction in the average escape time generated by an increasing in the noise intensity. For low noise intensities we have
\begin{equation}
	\left|\frac{d\bar{T}_l}{d\xi}\right| < \left|\frac{d\bar{T}_h}{d\xi}\right|.
\end{equation}

The previous equation implies that the algebraic law $\bar{T}_h$ fails in 
its predictions for low noise intensities because it is overestimating the effect of noise on the average escape time. Hence, it is clear that some factor is reducing the decrease of $\bar{T}$ for low noise intensities, in comparison with the observed behavior for $\xi\gg 0$.

To generalize the above result, we have computed the dependence of the average escape time on noise in the Barbanis potential, given by
	\begin{equation}
       {\cal{H}}=\frac{1}{2}(p_x^2+p_y^2)+\frac{1}{2}(x^2+y^2)-xy^2.
	\end{equation}
	 
	In this different Hamiltonian we have obtained qualitatively identical results to the ones that we have shown in Fig.~\ref{fig5}. 

It is important to notice that this scenario is radically different from the one that appears in noisy open Hamiltonian systems. When the system is open, the limit of Eq.~(\ref{lim1}) does not apply since for very low noise intensities the escape time tends to the finite value of the noiseless system. In this situation, the most characteristic effect of noise is the noise-enhanced trapping \cite{Altmann, NietoNET}, a phenomenon in which for 
a critical value of the noise intensity the average escape time increases. The main reason of this is that some unusual trajectories can decrease their energy under the threshold value $E_e$ and then describe very long transients. In fully chaotic systems the noise-enhanced trapping generates a global maximum in the average escape time evolution, which means that the average escape time of the noisy system exceeds that of the noiseless case. In mixed-phase-space systems the average escape time for $\xi=0$ is very high due to the stickiness, so instead of a global maximum 
we find a relative maximum that appears after a decrease generated by the 
gradual destruction due to the effect of the KAM islands. As we could observe in Fig.~\ref{fig5}, when the system is closed no maximum appears. This means that in this situation the noise-enhanced trapping does not influence the evolution of $\bar{T}$, that here is characterized by two monotonically decreasing regimes.

\section{Probability of escapes and average escape time distributions}\label{PE}
Since the noise-enhanced trapping is not responsible of the change in the 
scaling law of the average escape time, we should focus our attention on the other main factor that could affect the escape dynamics: the stickiness of the KAM islands.

The regular motion in Hamiltonian systems can be illustrated in many ways, such as through representations in the physical space or in Poincaré 
sections. However, when the objective is to quantify the size or to ascertain the existence of KAM islands, one of the most clear portraits is the 
computation of the KAM regions in the exit basins, defined as the set of initial conditions that do not escape when the time goes to infinity. In the presence of noise, two identical initial conditions can escape through different exits in different times. For this reason the usual exit basins representation does not offer relevant information. In particular, after a unique simulation of the exit basins we simply observe that they appear smeared by dispersed KAM regions and its boundary becomes blurred. Nevertheless, by adopting a probabilistic point of view, it is possible to construct basins of probability in noisy systems. Once a grid of initial conditions in the region of interest is defined, we associate a number between $0$ and $1$ to each of them depending on the probability of some asymptotic behavior. In general, we are interested in either the probability of some of the exits or in the probability of escape. The later allows us to check the existence of stickiness, since in the fully chaotic system the probability of escape should be uniform, while under the influence of the stickiness some regions with lower probability surrounding the KAM islands will appear.

\begin{figure}[htp]
	\centering		
	\includegraphics[width=1\textwidth,clip]{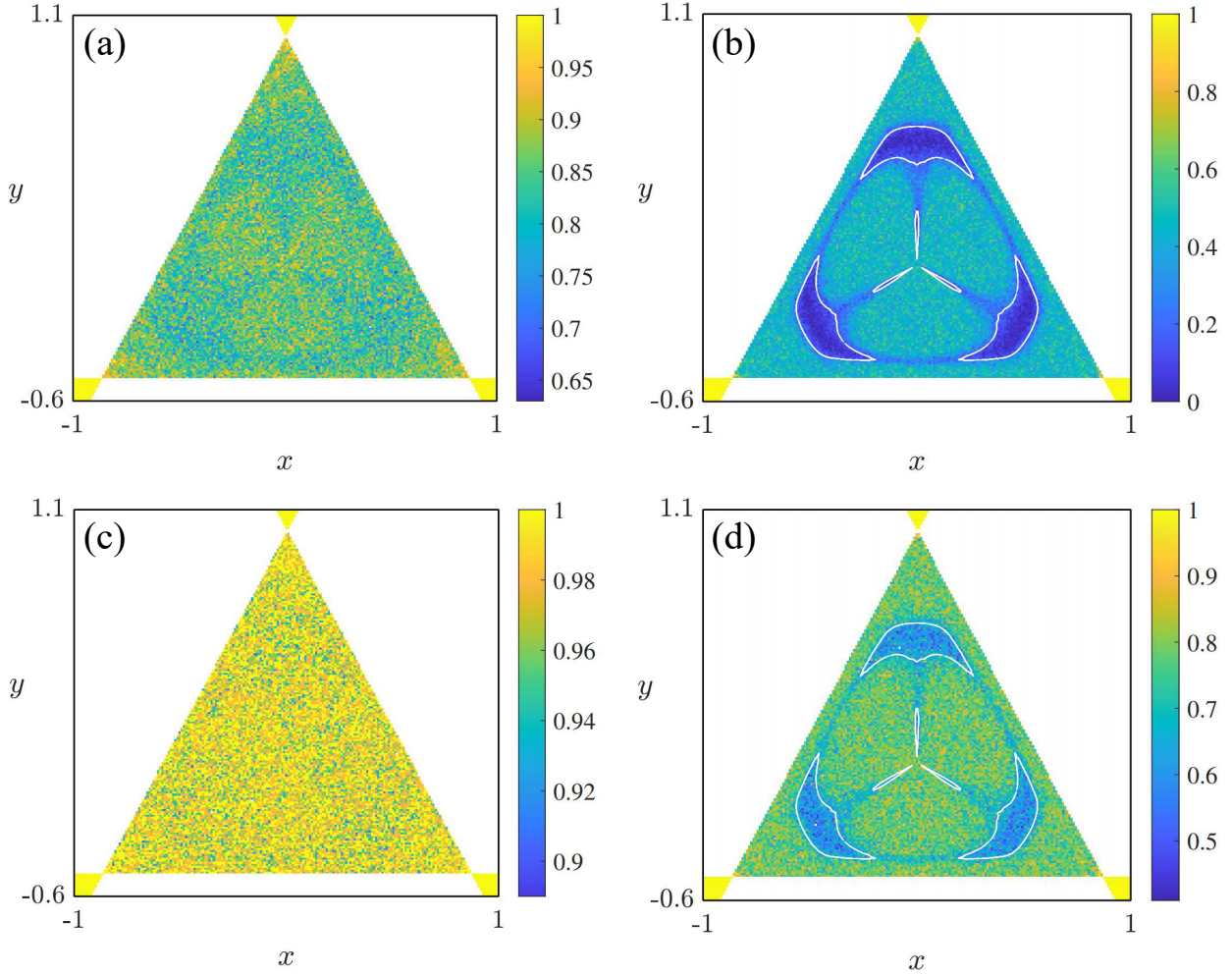}	
	\caption{Basins of probability showing for the Hénon-Heiles system with $E=1/6$ and noise intensities $\xi=10^{-5}$ (a,c) and $\xi=10^{-7}$ (b,d). The maximum integration time is $t_a=t_b=1000$ and $t_c=t_d=5000$. The basins have a $200\times 200$ resolution and every initial condition has been launched $200$ times before calculating the probability of escape. Hot and cold colors indicate high and low probability of escape, respectively. The white curves represented in panels (b) and (d) correspond to the boundary between the KAM islands and the chaotic sea.}
	\label{Fig6}
\end{figure}

To carry out the numerical simulation of the probability basins, we have computed $200$ exit basins in the physical space with a resolution $200\times200$ and calculated the probability of escape of every initial condition. We have done this for two different noise intensities $\xi=10^{-5}>\xi_h$ and $\xi=10^{-7}<\xi_l$. Since the probability of all the initial conditions tends to $1$ as the maximum integration time $t_{max}$ increases, the result could be unclear for very low or high values of $t_{max}$. In order to compare the result for different times, we have chosen $t_{max}=1000$ and $t_{max}=5000$. The result is shown in Fig.~\ref{Fig6}, where a different probability distribution is observed for $\xi<\xi_l$ [panels (b) and (d)] and  $\xi>\xi_h$ [panels (a) and (c)]. For noise intensities under the threshold $\xi_l$, the stickiness manifests by decreasing the probability of the escape in the neighborhood of the KAM regions of the conservative system. To visualize these regions, we represent in white the last KAM curve in Fig.~\ref{Fig6} [panels (b) and (d)]. This curve is the outer limit of the KAM island and beyond it we find the chaotic domain. For $t_{max}=1000$ [panel (b)] the initial conditions that start inside a KAM island cannot escape due to the effects of noise. When the maximum integration time is increased until $t_{max}=5000$ [panel (d)] the same effect is observed on the distribution, but in this case the minimum probability of escape is close to $0.4$. For the noise intensity over the threshold, all the particles have the chance of escaping after $t_{max}=1000$ [panel (a)] and an almost negligible stickiness is observed. If the maximum time is increased until $t_{max}=5000$ [panel (c)] the probability distribution is almost uniform, ranging 
between $0.9$ and $1$.

\begin{figure}[htp]
	\centering
	\includegraphics[width=1\textwidth,clip]{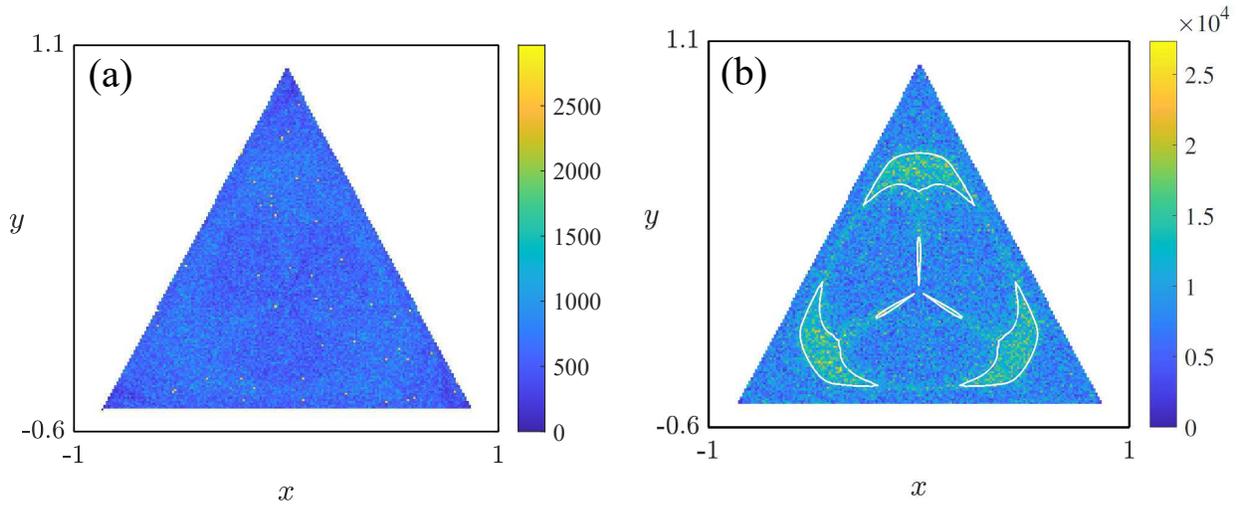}
	\caption{Average escape time distribution for the Hénon-Heiles system with $E=1/6$ and noise intensities $\xi=10^{-5}$ (a) and $\xi=10^{-7}$ (b). The figures have a $200\times 200$ resolution and every initial condition has been launched $200$ times before calculating the average escape time. We have used a very high maximum integration time $t=10^{5}$ in order to guarantee that all the trajectories have escaped the region at the end of the numerical simulation. Hot and cold colors indicate high and low average escape times, respectively. The white curves represented in panel (b) correspond to the boundary between the KAM islands and the chaotic sea.}
	\label{Fig7}
\end{figure}

The evidence that we have shown in Fig.~\ref{Fig6} confirms that the stickiness influences the escape dynamics for noise intensities $\xi<\xi_l$, while it is almost imperceptible for $\xi>\xi_h$. However, there is another important evidence whose trace does not disappear when increasing $t_{max}$ and it is based on the average escape time distributions. In the absence of stickiness that could retain the particles, the average escape time distribution should not exhibit compact regions with high escape times, located in the vicinity of the KAM islands. On the other hand, if the stickiness influences the noisy system, two regions with different ranges of average escape times are expected. To show this, we represent in Fig.~\ref{Fig7} the average escape time distribution in the physical space for again noise intensities (a) $\xi=10^{-5}$ and (b) $\xi=10^{-7}$, that are over $\xi_h$ and under $\xi_l$, respectively. 
The result is as predicted, and shows an absolutely different distribution 
for high and low noise intensities. For low noise intensities the initial conditions launched from a KAM region (see white curves in Fig.~\ref{Fig7}) describe in average longer transients ($\bar{T}>15000$) than the chaotic initial conditions ($\bar{T}<15000$). On the 
other hand, for high noise intensities most of the particles escape in short times ($\bar{T}<750$), while some dispersed initial conditions reach maximum average escape times close to ($\bar{T}=3000$).

\begin{figure}[htp]
	\centering
	\includegraphics[width=0.55\textwidth,clip]{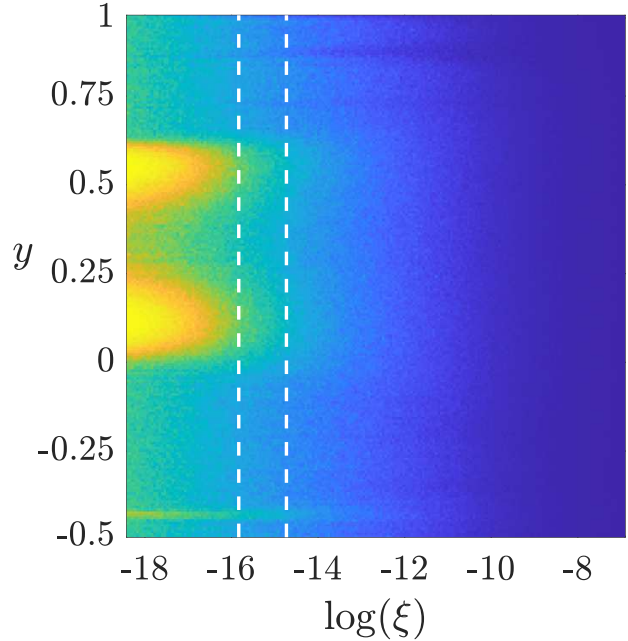}
	\caption{Color-code map showing the average escape time of initial conditions in the segment $x=0$ of the physical space for different values of the natural logarithm of noise intensity. The white dashed lines are located at the critical values of the noise $\xi_l=1.5\times10^{-7}$ and $\xi_h=4\times10^{-7}$. Hot and cold colors indicate high and low average escape times, respectively. }
	\label{Fig8}
\end{figure}

To get a more general view of the gradual reduction of the stickiness, we represent in a color-code map the average escape time along the segment $x=0$ in function of the noise intensity in Fig.~\ref{Fig8}. To generate this figure we have chosen $200$ equally spaced initial conditions in the range $y_0\in[-0.5,1]$ and calculated the average escape 
time for $200$ values of the noise $\xi\in[10^{-8},10^{-3}]$. We use a logarithmic scale in the axis of the noise in order to appreciate the qualitative changes better. For values $\xi<\xi_l$, three regions of high average escape time are visible.These regions can be observed in both Fig.~\ref{Fig6} and Fig.~\ref{Fig7} and correspond to the horizontal thin KAM region close to $y=-0.5$, the thin vertical KAM region starting close to $y=0$ and the wide KAM region centered approximately in $y=0.5$. Once the noise intensity surpasses $\xi_h$ these regions disappear, indicating the end of the stickiness. Inside the transition region $\xi\in[\xi_l,\xi_h]$, indicated in Fig.~\ref{Fig8} with vertical dashed lines, the stickiness affects the system, but it is almost negligible.

The previous results provided strong numerical evidence confirming that for low noise intensities the escape dynamics is still affected by the stickiness, while for high noise intensities it is negligible. In some systems the KAM islands exhibit a self-similar structure with typical size $l$. In these cases the critical noise can be predicted since $\xi_c \propto l$ \cite{Seoane09}. However, in usual Hamiltonian systems KAM islands of different size and topology coexist in the phase space, so their stickiness is also different and the same noise intensity affects them in a different way. In this scenario the destruction of the stickiness can be understood more clearly by changing the perspective from the noise to the energy fluctuations.

\section{Energy fluctuations}\label{EF}

The main reason explaining that trajectories that are in or close to a KAM island leave that region in short times is based on the energy fluctuations. The intensity of the trembling of the potential well depends on the intensity of the fluctuations of the energy, which is related with the noise intensity. In particular, the higher the noise, the larger the variance of the energy fluctuations and the probability of escape.

There are two mechanisms that allow the escape of these particles. One possibility is that the random fluctuations increase the energy of the system to values where the KAM structures are absent or their size is negligible. On the other hand, because of the noise effects some particles can travel over a long enough distance in the phase space, reaching fully chaotic regions and hence escaping in short times.

 \begin{figure}[htp]
 	\centering
 	\includegraphics[width=0.45\textwidth,clip]{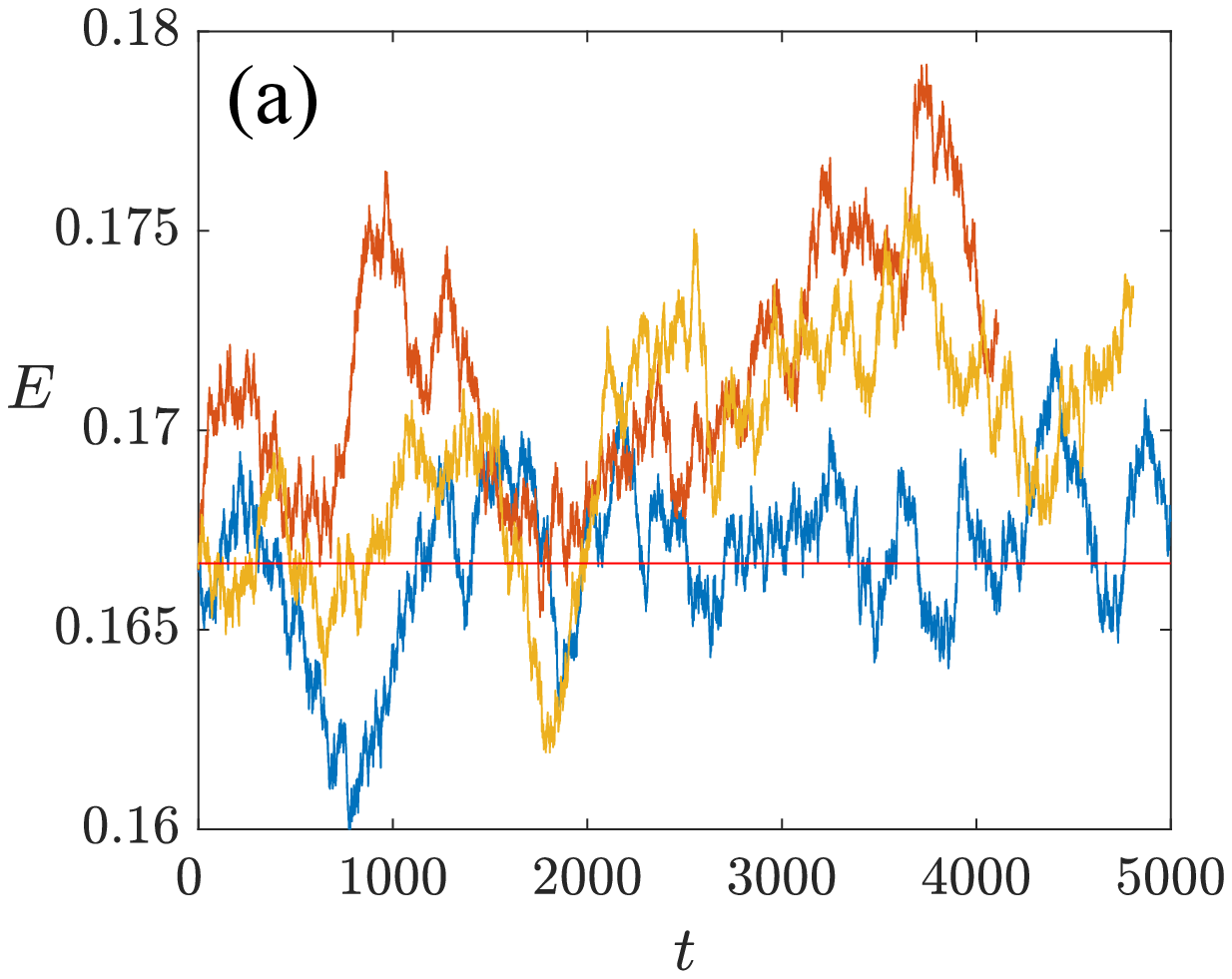}
 	\includegraphics[width=0.45\textwidth,clip]{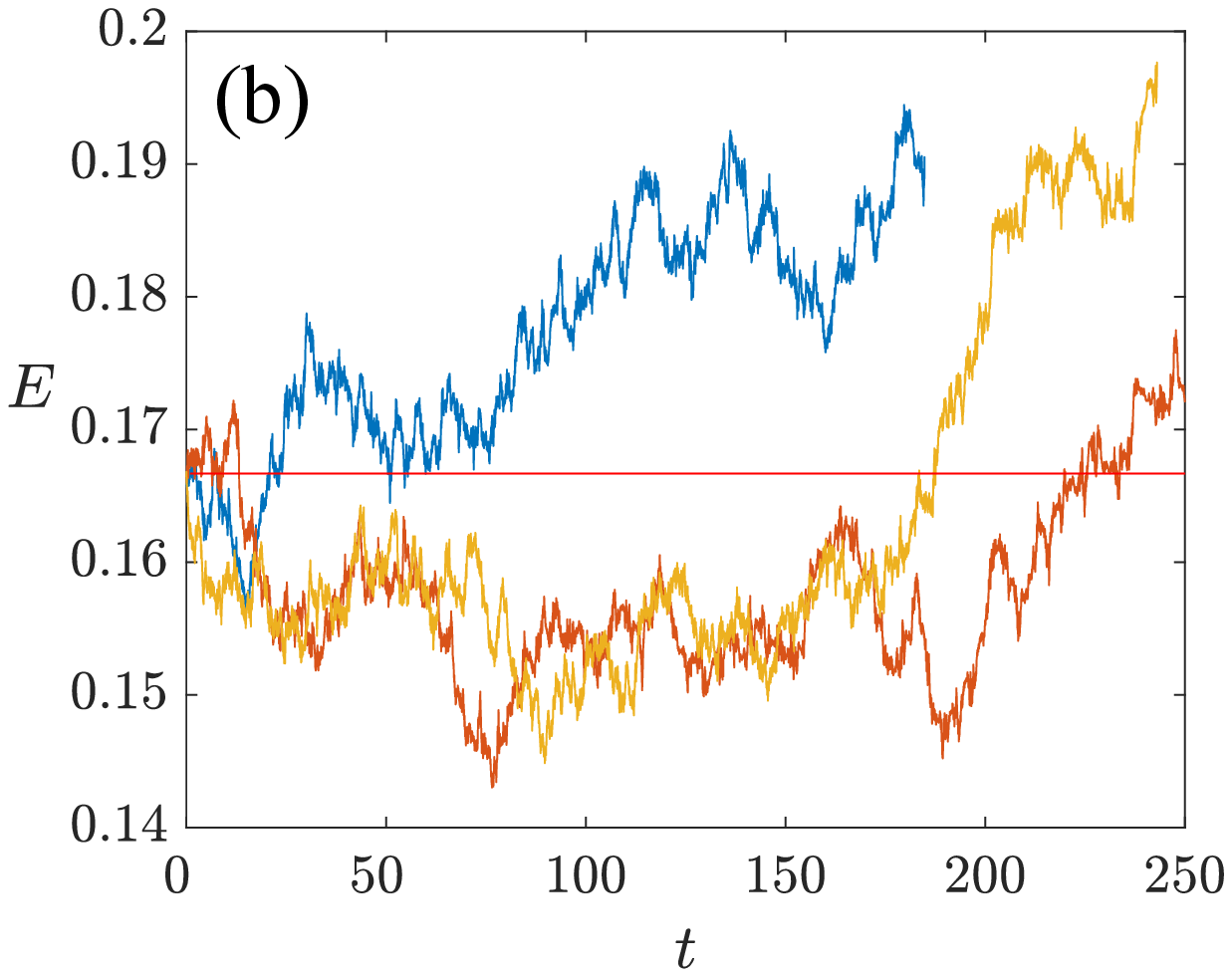}
 	\caption{Energy evolution of different launchings of the same initial condition in the presence of a noise intensity (a) $\xi=10^{-7}$ and (b) 
$\xi=10^{-5}$. In both cases the initial condition belongs to a KAM region in the noiseless system. The red lines indicate the initial energy $E=1/6$. }
 	\label{Fig9}
 \end{figure}

To visualize the different energy evolution depending on the noise intensity, we show in Fig.~\ref{Fig9} three simulations of a particle starting in 
a KAM island for (a) $\xi=10^{-7}<\xi_l$ and (b) $\xi=10^{-5}>\xi_h$. 
The weakly noisy trajectories cannot reach high energies after a long time $t\approx 5000$, and they move with energy values where the conservative dynamics is dominated by extensive KAM islands. On the contrary, the trajectories affected by a noise intensity over the threshold exhibit huge fluctuations in short times $t<250$, reaching energy values close to $E=0.2$ for which KAM islands are insignificantly small \cite{NietoND}.

\begin{figure}[htp]
	\centering
	\includegraphics[width=0.5\textwidth,clip]{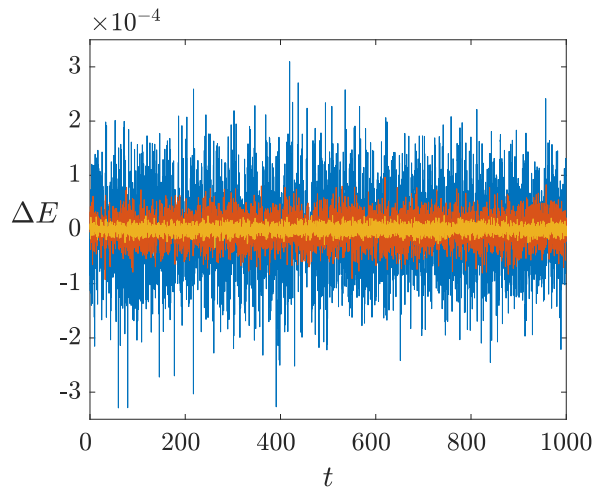}
	\caption{Variation of the energy in every integration step of three noisy trajectories of the Hénon-Heiles with $E=1/6$. The noise intensities are $\xi=10^{-5}$ (blue), $\xi=10^{-6}$ (yellow) and $\xi=10^{-7}$ (orange). }
	\label{Fig11}
\end{figure}

The representation of the evolution of the energy gives an account of the 
usual range of energy levels that a particle can reach in function of the 
noise intensity. This evolution is generated by a chain of energy variations $\Delta E$ that are introduced in every integration step $h$. The variance of $\Delta E$ is proportional to the variance of the noise intensity and it is, in essence, what conditions the probability of escape. In this sense, the complete destruction of the stickiness and hence the change in the scaling law of the average escape times, can be seen equivalently as a consequence of a critical noise or the $\Delta E$ distribution. To illustrate this, we represent the variation of the energy in every integration step for three trajectories affected by different noise intensities $\xi=10^{-5}$ (blue), $\xi=10^{-6}$ (yellow) and $\xi=10^{-7}$ (orange) in Fig.~\ref{Fig11}. It is difficult to imagine the effect that a noise intensity has in a system by just looking to its value. However, Fig.~\ref{Fig11} is a portrait that shows the perturbations in the energy that the quasiperiodic orbits can or cannot endure. In the lower energy case, the trajectory that starts in a KAM region still has some probability of falling again inside the KAM torus after the noise has perturbed its energy. This can continue during a long transient after which a set of unfavorable fluctuations will generate the escape. The situation in the higher energy case is radically different. The magnitude of the fluctuation is so huge that the trajectory can jump out of the KAM torus in a short time.

It is interesting to notice that the variation in the energy does not appear exclusively in noisy systems, but in all the perturbed Hamiltonian processes such as forced and dissipative chaotic scattering problems. In these examples, that are characterized by completely different phenomena, the energy variation plays a relevant role in the destruction of the KAM islands.

\section{Conclusions}\label{Conc}
In summary, our research reveals that a noise-activated escape occurs in closed Hamiltonian systems. The fundamental effect of noise is to generate random fluctuations in the energy that allow the particles to reach energies over the threshold $E_e$ and escape. In this situation the system can be studied by using the usual tools of chaotic scattering. The evolution of the average escape time exhibits a transition region that separates two different regimes, characterized by different algebraic scaling laws. In particular, if the noise is decreased starting from high values $\xi>\xi_h$, the evolution of $\bar{T}$ is explained by an algebraic scaling law $\bar{T}_h(\xi)$. On the other hand, for low noise intensities $\xi<\xi_l$, the evolution is governed by a different algebraic law $\bar{T}_l(\xi)$. Between these critical thresholds a smooth transition appears, and it is characterized by an almost negligible stickiness. We have uncovered that the key reason under this change in the regime is the destruction of the stickiness of the KAM islands, that slows down the decrease of the average escape time in comparison with the highly noisy behavior. To show this, we have provided strong numerical evidence based on the probability of escape and average escape time distributions in the physical space. In both cases, for $\xi<\xi_l$ regions with low probability of escape or high average escape time appear, surrounding the regions occupied by the KAM islands in the noiseless system. On the contrary, for noise intensities $\xi>\xi_h$ an almost uniform distribution is observed, indicating the complete destruction of the stickiness.

Under the influence of noise, both open and closed Hamiltonian systems are examples of chaotic scattering. Moreover, by simply analyzing a noisy temporal series with an unknown initial condition we cannot guess if it comes from an open or a closed Hamiltonian. This fact brings the possibility that certain physical systems can be modeled through noisy closed rather than open Hamiltonian systems. Since the escape dynamics is different in both cases, using one of the two models can be more appropriate depending on the behavior of the system.

We hope that this research could contribute to a deeper understanding of noisy Hamiltonian systems. We expect potential applications in several fields of science where chaotic scattering problems arise. Some examples are the motion of stars in galaxies \cite{Navarro,Contopoulos02} and restricted cases of the three-body problem \cite{Assis,Bernal20}. Moreover, the effect of weak noise has important consequences in the phase mixing of chaotic orbits, that has implications in a wide variety of physical situations \cite{Kandrup}.

 \section*{ACKNOWLEDGMENTS}
This work has been financially supported by the Spanish State Research Agency (AEI) and the European Regional Development Fund (ERDF) under Project No. PID2019-105554GB-I00.

\end{document}